# Exotic Pressure-Driven Band Gap Widening in Carbon Chain-Filled KFI Zeolite and Its Pathway to High-Pressure Semiconducting Electronics and High-Temperature Superconductivity

[1]C.T.Wat, [1]K.C.Lam, [1]W.Y.Chan, [1]C.P.Chau, [1]S.P.Ng, [1]W.K.Loh, [2]L.Y.F.Lam, [2]X.Hu, [1]C.H.Wong*

[1]*Division of Science, Engineering, and Health Studies, The School of Professional Education and Executive Development, The Hong Kong Polytechnic University, Hong Kong, China*

[2]*Department of Chemical and Biological Engineering, The Hong Kong University of Science and Technology, Hong Kong, China*

Email: roy.wong@cpce-polyu.edu.hk

**Abstract:**

Semiconducting devices face persistent challenges in operating at high pressure, as the band theory predicts that materials transition to a more metallic state under compression. However, our findings with carbon chains in KFI substrates reveal a conditional deviation from this norm. We not only witness the transition from polyyne (semiconductor) to cumulene (metal) at medium pressure, but we also observe an unexpected re-entrance of the polyyne at high pressures, where the band gap in the polyyne increases with pressure. In addition, the synthesis of long cumulene chains has posed a longstanding challenge in the quest for high-temperature organic superconductivity. We have identified critical conditions for synthesizing extended cumulene chains within zeolite frameworks, highlighting the interplay between unconventional charge density waves and significant torsions. The KFI zeolite facilitates the formation of carbon chains exceeding 5,000 atoms, in stark contrast to around 100 other zeolites that are limited to ~10 atoms. The cumulene@KFI system demonstrates a superconducting transition temperature reaching ~62 K, surpassing the highest reported values for bulk iron-based superconductors. This interplay between carbon structures and superconductivity not only advances our understanding of charge density waves but also heralds a new era in the study of novel applications

## 1. Introduction:

Carbon-based electronics has become a prominent field of research thanks to the unique properties of carbon [1], including its topological states of matter, semiconducting abilities, and superconductivity, among others. However, most materials, regardless of whether they are carbon-based or not, transition to a metallic state under high pressure. This phenomenon presents a longstanding challenge for the development of high-pressure semiconducting devices, as maintaining or widening a band gap value under such conditions is tough based on the band theory. On the other hand, the superconducting transition temperatures of organic superconductors have consistently lagged that of other unconventional superconductors [2-3]. Although carbon chains have emerged as a promising candidate to challenge this notion, primarily due to the presence of a van Hove singularity in the electronic density of states (DOS) and a high Debye temperature of approximately 3000 K [4-6], carbon chains predominantly exist in the polyyne phase. This phase is characterized by alternating single and triple bonds, which results in a semiconductor with a large band gap that impedes Cooper pair formation. The realization of the metallic cumulene phase in carbon chain, characterized by consecutive double bonds [9,10], could address this issue. However, the formation of a long cumulene chain presents a significant unresolved challenge, both experimentally and theoretically. Although studies on a long cumulene chain are present in worldwide theoretical literature, analysis is limited by the manual specification of double bonds instead of forming naturally. The favorable conditions for a long cumulene remain an open question, as does the pursuit of high-temperature organic superconductivity. The cumulene chain needs to be sufficiently long; otherwise, it cannot fully leverage the advantages of Von-Hove singularities in the electronic density of states for building strong Cooper pairs.

Triggering charge modulation along a cumulene chain could produce effects akin to the periodic charge modulation observed in polyyne. Creating a charge density wave at the Fermi surface poses a complex challenge in carbon materials, as such phenomena typically need a strong electron-electron interactions, e.g. heavy-element composites [11,12]. However, encapsulating carbon chains within boron-nitride nanotubes (BNT) could induce an unconventional charge density wave due to the periodic electric field from the host, while also alleviating magnetic frustration [13]. However, the unconventional charge density wave does not transform the internal carbon chain into the cumulene phase in the BNT host, indicating that while this wave may be necessary, its magnitude is another obstacle to achieving cumulene.

A similar host system involving dissimilar atoms could be zeolites, which have pores acting as nanoreactors. These structures could provide favorable van der Waals forces and heterogeneous interactions to influence carbon chains, potentially leading to the formation of cumulene chains. From manufacturing point of view, carbon chains are often limited to fewer than 10 atoms [7] unless a suitable host is identified such as carbon nanotubes [8]. Determining which zeolite systems can facilitate the formation of such long carbon chains is another open question.

Historically, predictions about the length of carbon chains have primarily relied on empirical experimentation. However, in 2017, C.H. Wong et al. introduced an LCC (Linear Carbon Chain) array model to predict the relationship between the cut-off length of carbon chain arrays and their magnetic properties [14]. Our research revealed that a lateral distance of 0.5 nm significantly enhances the likelihood of forming a parallel array of carbon chains, with expected lengths around 50 atoms [14], where a finding that was experimentally validated in 2018 [15]. In 2023, C.H.Wong et al further refined the LCC array model by replacing the neighboring chains with CNT [16]. The LCC@CNT model demonstrated that a stable ~5700-atom-long carbon chain could exist within a (6,4) CNT, with experimental findings confirming a length record of ~6400 atoms [16]. Inspired by this progress, we are now developing the LCC@Zeolite model to predict which zeolites could enable the formation of long carbon chains. Then we will illustrate the potential of these chains to exist in the cumulene phase while exhibiting both unconventional charge density waves and high-temperature superconductivity [17].

## 2. Methods

The repeated units of the carbon chain filled into over 100 zeolite substrates (supplementary materials) are built in the ab-initio software (CASTEP) at 0K under the GGA-PBE functional [19], respectively. The Density Mixing approach is employed, with an SCF cycle set to 5000. An ultrasoft potential is utilized, and the Finite Displacement Method [20] is used to estimate phonons. Geometric optimization of atomic coordinates is performed again under different pressures to examine the effects of compressive strain. However, since the carbon chain is not infinitely long in reality, it is essential to account for size effects. These C-

chains@Zeolite substrates will be subjected to the Monte Carlo simulation to predict the cut-off chain length at 300K.

By replacing the host to zeolite substrate in the Monte Carlo simulation [13,16], we build C-chain@zeolite model where the Hamiltonian is,

$$H = e^{-T/T_{bj}} \left( \sum_{n=1,3,5}^{N} |E_{n,j} - E_1| e^{-\frac{\ell_n - \ell_{n,j}^{eq}}{0.5\ell_{n,j}^{eq}}} + \sum_{n=2,4,6}^{N} |E_{n,j} - E_3| e^{-\frac{\ell_n - \ell_{n,j}^{eq}}{0.5\ell_{n,j}^{eq}}} \right) + e^{-T/T_{bj}} \sum_{n=1,2,3}^{N} J_A e^{\frac{\ell_n - \ell_{n,j}^{eq}}{0.5\ell_{n,j}^{eq}}} (\cos\theta + 1)^2$$

$$-4\varphi \sum_{\phi=0}^{2\pi} \sum_{n=1}^{N} \left[ \left(\frac{\sigma}{r}\right)^6 - \left(\frac{\sigma}{r}\right)^{12} \right]$$

This model defines multiple structural parameters: the distance between adjacent bonds ($\ell_n$) and the lateral separation between the linear carbon chain (LCC) and the enclosed atoms in zeolite [13,16]. Here, j encodes the bond type and its associated energy: 1 for a single bond ($E_1$ = 348 kJ/mol, $\ell_{n,1}^{eq}$ = 154 pm), 2 for a double bond ($E_2$ = 614 kJ/mol, $\ell_{n,2}^{eq}$ = 134 pm), and 3 for a triple bond ($E_3$ = 839 kJ/mol, $\ell_{n,3}^{eq}$ = 120 pm) [13,16]. The index $n$ specifies the position of carbon atom in the chain. Thus, the notation $E_{n,j}$ denotes the bond energy for the bond between the $n$ th atom and its predecessor, the $(n-1)$ th atom. The characteristic dissociation temperature for a given bond type is derived as $T_{bj} = E_{n,j} / k_B$, where $k_B$ is the Boltzmann constant.

To evaluate the thermodynamic stability of the confined carbon chain, a chain-stability factor $e^{-\frac{\ell_n - \ell_{n,j}^{eq}}{0.5\ell_{n,j}^{eq}}}$ is introduced [13,16]. The host-chain interaction is explicitly modeled through a Van der Waals (VDW) potential term $E_{vdw} = -4\varphi \sum_{\phi=0}^{2\pi} \sum_{n=1}^{N} \left[ \left(\frac{\sigma}{r}\right)^6 - \left(\frac{\sigma}{r}\right)^{12} \right]$ along the lateral plane. This potential is parameterized by constants $\sigma$ and $\varphi$, which are derived from physical properties including isothermal compressibility and characteristic system length scales respectively [13,16]. Furthermore, an angular bending potential of $J_A \approx 600$ kJ/mol governs the energy penalty associated with deviations from linearity at the carbon pivot angles (*θ*) along the chain backbone [13,16].

The dynamic behavior of the encapsulated carbon chain is simulated using the Monte Carlo methodology. This approach iteratively samples the configurational space by allowing stochastic modifications to both the atomic coordinates and the covalent bond types within the carbon chain at finite temperature. Each Monte Carlo Step (MCS) is structured according to the following algorithm, adapted from the protocol [13,16].

1. **Selection:** A single atom within the carbon chain is randomly chosen.
2. **Energy Evaluation:** The total energy of the chain, defined by the Hamiltonian, is calculated for geometric configuration.
3. **Trial Move Generation:** A new trial state is proposed by introducing a random spatial displacement for the selected atom and/or stochastically altering its local bond type (see the supplementary materials – Part A and Part B)
4. **Trial Energy Calculation:** The Hamiltonian is recalculated for this proposed trial configuration.
5. **Criterion:** The trial state is accepted if it results in a lower total energy. Increase in the trial states may be also accepted by comparing a probability between 0 and 1 with the Boltzmann factor, $e^{\frac{-\Delta E}{k_B T}}$, where ΔE is the energy difference, $k_B$ is Boltzmann's constant, and *T* is the temperature. Otherwise, the system reverts to its previous state. The Monte Carlo simulation proceeds iteratively until the system reaches thermodynamic equilibrium, which is operationally defined as the point where the total energy as a function of Monte Carlo Steps (MCS) plateaus. Based on established convergence from prior work [13,16], equilibrium is typically achieved after ~150,000 MCS.

The search for a long carbon chain in the cumulene phase within zeolite substrates will continue. Concurrently, the properties of electrons and phonons will be evaluated to predict superconductivity. The T$_c$, based on electron-phonon coupling $\lambda \sim 2\int \alpha^2 \frac{F(\omega)}{\omega} d\omega$, will be calculated using the McMillan $T_c$ formula [21], where α is the electron-phonon scattering matrix and F(ω) is the phonon density of states as a function of vibrational frequencies in the form of the Eliashberg function [21]. If the electron-phonon coupling λ exceeds 1, we typically apply a renormalization approach by dividing by 1+λ [22]. The same renormalization factor (1+λ) can be applied to the pseudopotential μ [22].

**3. Results and Discussions.**
**3.1. Characteristics for the Growth of Long C Chains in Zeolite**

We present the lattice parameters of over 100 zeolites in the supplementary materials (Part C). After evaluating the stability of over 100 different C-chains@zeolite substrates (supplementary materials – Part C), we found that the KFI zeolite is a promising option for stabilizing the carbon chain up to ~5500 atoms in length at ambient condition as shown in Figure 1, while all other zeolite substrates fail to stabilize chains longer than 10 atoms. Our MCS step is sufficiently long to relax the C-chains within KFI Zeolite, as the chain length is slightly shorter than that in (6,4)CNT. In the case of C-chain@(6,4)CNT, the MCS is only 150,000 [13]. Therefore, our chosen MCS value for this project is adequate. Due to the stable long-chain characteristics, the formation of free radical electrons is not noticeable in the internal chain.

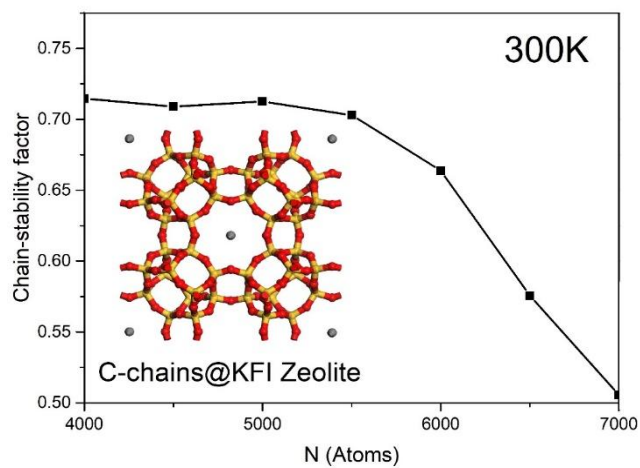

Figure 1: The chain stability factor of the C-chains inside KFI Zeolite in the absence of external pressure. The cut-off length of the internal chains is ~5500-atom-long. The inset figure shows the side view of the repeated unit. The grey balls are carbon. The yellow and red balls build the KFI Zeolite. The lateral chain-to-pore distance is ~0.54nm. The lateral chain-to-chain distance is ~1.3nm.

Finding a suitable host to grow C chains of this length is challenging, whether through experimental discovery or theoretical proposals. The unique finding in KFI zeolite raises our curiosity about the reason for the long-chain fabrication. KFI zeolite shares the same lattice parameters and crystalline angles in the x-, y- and z-axis. In the supplementary materials (Part C), we have identified eight zeolites (green color) exhibiting similar behavior. However, CHA zeolite, with an angle of 94.2 degrees, is eliminated from the second shortlist. This leaves us with seven candidates (see the supplementary materials – Part D). A record-breaking chain length of 6400 carbon atoms has been achieved in a (6,4)CNT, which has a tube radius of ~0.37 nm. We have observed a similar feature in KFI zeolite, where its pore radius is 0.35 nm, closely matching the radius of the (6,4)CNT. This correlation elucidates the fabrication of long carbon chains in KFI zeolite.

Three reasons are summarized to explain the development of this long chain. First, the radius of the (6,4)CNT and the radius of the KFI zeolite pore are comparable. Secondly, the lattice parameters of KFI zeolite are uniform along the x, y, and z axes, Thirdly, the crystalline angles in the lattice are all 90 degrees. This symmetrical nature of the pore helps prevent asymmetric lattice expansion, which could impede chain growth. Notably, only C-

chains@KFI substrate among the ~100 zeolites examined exhibits all three of these characteristics.

**3.2: Unconventional Effects of Stress on the Band Gap**

To actuate the formation of superconductivity, we search for methodologies to convert it to cumulene and therefore, we estimate the band gaps of the C-chains@KFI Zeolite. The cut-off chain length of 5500 atoms is large enough to be regarded as a bulk unit in the ab-initio calculations. Before the incorporation of carbon chains, pristine KFI zeolite displays a large band gap above 4 eV. At first glance, reducing the band gap to zero appears technically challenging, as significant band-gap modulation in semiconductors is difficult. Surprisingly, after filling the KFI zeolite with carbon chains, the band gap of the composite is drastically reduced by an order of magnitude to 0.47 eV. This surprising 90% reduction in band gap raises the possibility of transitioning into a metallic phase (cumulene) with the application of pressure. More exciting developments are emerging. Typically, applying pressure brings atoms closer together, which tends to reduce the band gap based on the band theory. However, when a compressive strain of up to 4% is applied to the composite, it unexpectedly widens the band gap to 0.83 eV in Figure 2a. Fascinatingly, at approximately 5% strain, the material transitions into a metallic phase, marked by a vanishing band gap. Concurrently, the carbon chains change into the cumulene phase. Even more surprisingly, we expect the metallic cumulene to persist with compressive strains greater than 5% by symmetrically reducing the bond lengths. Yet, instead, the composite shifts back to a semiconductor state, accompanied by the re-entrance of the polyyne phase.

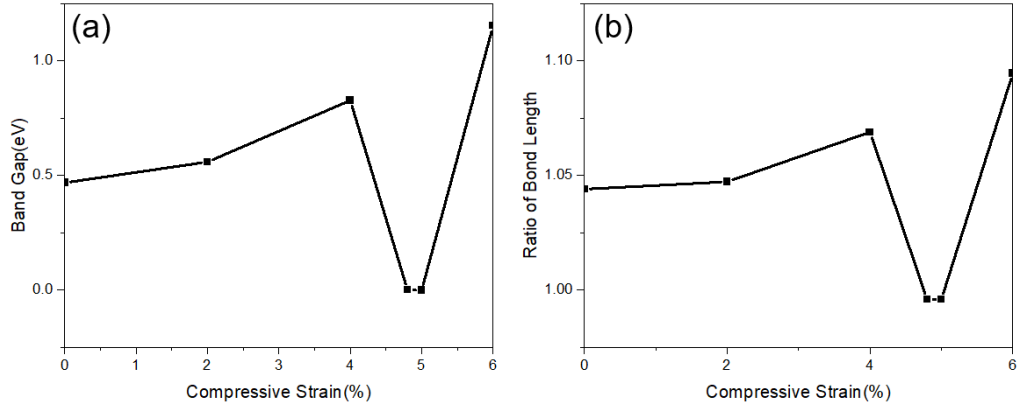

Figure 2: (a) The band gap of the C chains in the KFI substrate. Applying compressive strain up to 4% leads to an adverse increase in the band gap. Around 5% strain, the band gap decreases to zero, promoting the formation of a cumulene structure. Surprisingly, applying a compressive strain greater than 5% causes the material to revert back to a polyyne phase. (b) The ratio of single to triple bond lengths exhibits an unusual increase with pressure, except that this ratio reaches 1 at approximately 5% compressive strain owing to the phase transition.

This series of results highlights a highly unusual and complex pressure effect on the electronic properties of C chains, offering new insights into solving a long-standing puzzle regarding how to increase band gap values in both low- and high-pressure electronics. For the sake of convenience, we applied the same compressive strain to both the lateral and longitudinal axes, as we observed that the band gap of C-chains@KFI Zeolite is not very sensitive to lateral pressure. The exotic relationship between the band gap and pressure is consistent across other DFT setups, such as PW91, LDA, norm-conserving potentials, and various SCF cycles, etc. Replacing CASTEP with Dmol3 software yields similar band gap values.

The inverse proportion between band gap and pressure can be due to the exceptionally strong covalent bonds within the C chain [23] and the internal interaction from the zeolite. Unlike many conventional semiconductors, where applying pressure typically reduces the band gap, the C chain (recognized as the strongest material known) displays unique characteristics. While the average atomic separation of C chains in KFI Zeolite decreases under external pressure, the carbon atoms preserve their preferred alternating single and triple bond lengths. The strong covalent bonds in the C chains allow them to accommodate a larger bond-length ratio in response to pressure, rather than undergoing a transformation to a more metallic phase, as illustrated in Figure 2b. In contrast, other semiconductors, particularly those with optical modes, are rarely reported the same feature. These materials typically possess weaker bonding, which lacks the covalent strength needed to support the optical modes while average atomic distances diminish. As a result, pressure tends to drive these ordinary semiconductors towards a more metallic state, leading to a decrease in the band gap.

**3.3: Transition from Polyyne to Cumulene Induced by Charge Density Waves**

The unexpected transition from polyyne to cumulene within KFI Zeolite at the medium pressure (compressive strain ~5%) reveals a fascinating phenomenon, as shown in Figure 3a. Contrary to expectations, the transition to the cumulene phase may forfeit the benefit of periodic charge modulation among the shared electron clouds. Yet, intriguingly, Figure 3b shows that the charge density waves occur in the Fermi surface of the cumulene chain, defying conventional predictions. These charge density waves do more than just exist. They replicate the periodic charge modulation characteristic [16] of the polyyne phase, allowing the C chain to indirectly reclaim this favorable feature even while in the cumulene phase. This unexpected behavior not only stabilizes the cumulene phase but also challenges our understanding of charge dynamics in monoatomic materials. Another compelling reason for the formation of the cumulene phase is that, as shown in Figure 2b, the ratio of single to triple bonds approaches 1. This condition makes it challenging to stabilize the polyyne phase. By transitioning to the cumulene phase, the system can benefit from having uniform consecutive bond lengths. More interestingly, we observe significant torsional effects in the cumulene chain, with torsional (or twisted) angles reaching as large as 90 degrees in Figure 3a. This exceeds any previous understanding of carbon science, especially considering that torsioning the strongest material, the carbon chain, by 90 degrees is phenomenal due solely to the Van der Waals forces of the zeolite.

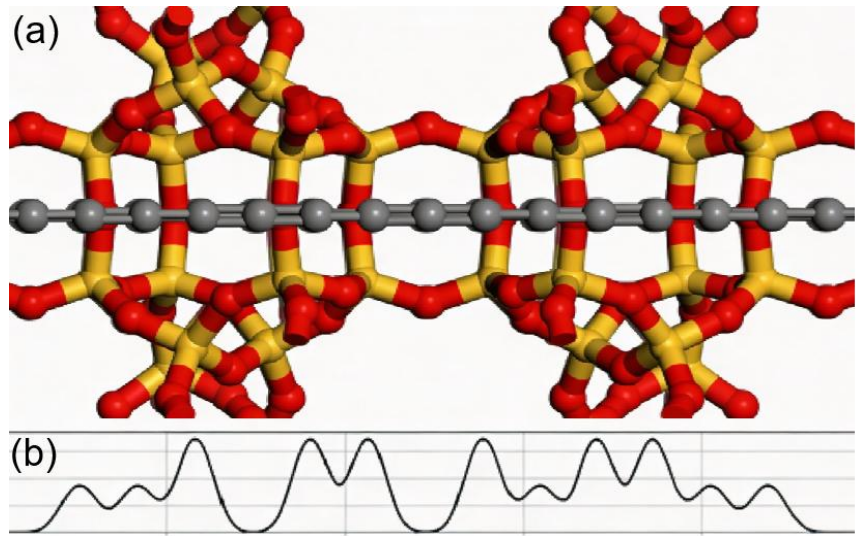

Figure 3: (a) Zoomed-in view of the cumulene chain within KFI Zeolite's, repeated unit under approximately 5% compressive strain. The naturally formed torsional angle of the internal cumulene chain (grey balls) is as large as 90 degrees by the application of Van der Waals force only. (b) Increase in electronic DOS at the Fermi level of the internal cumulene along the chain axis, expressed in arbitrary units. Charge density waves with varying amplitudes and wavelengths are superimposed, with a maximum change in amplitude of approximately 4%.

**3.4: Investigation of Superconductivity in Cumulene within KFI Zeolite**

The pursuit of superconductivity in carbon chains has faced challenges due to the presence of a semiconducting gap. However, with this band gap vanishing in C-chains@KFI zeolite at the medium pressure, we are compelled to investigate how far the superconducting transition temperature ($T_c$) could rise for cumulene within this substrate. Our calculated Debye temperature for isolated cumulene chains measures an impressive 2700 K [6]. Our findings reveal that while the introduction of torsional angles in the cumulene chain slightly reduces the Debye temperature by about 10%, it still remains remarkably high at approximately 2500 K. This elevated Debye temperature could robustly support Bardeen-Cooper-Schrieffer (BCS) pairing [4].

Our previous experimental analyses suggest that the substrate does not alter the superconducting transition temperature of nanowires [24]. To streamline our computational approach, we eliminate the KFI substrate in the ab-initio calculations, while maintaining fixed atomic coordinates and torsional angles of the cumulene chains for bare electron-phonon coupling assessments. The BCS framework appears valid without the complexities of spin-orbit coupling, anisotropic Fermi surfaces, or magnetic fluctuations in the composite system [4]. Despite the existence of a charge density wave, its modulation is minimal, allowing us to approximate the mean-field electronic DOS of the chain. In principle, we could apply the linear separation of variables method concerning the charge density wave to examine its effects on electron-phonon coupling, as successfully used in iron-based and cuprate superconductors [25-29], to refine our calculations. However, the impact is considerably smaller here, given that the charge density wave amplitude is significantly less than in the unconventional superconductors. Furthermore, here the charge density wave acts primarily to eliminate the semiconducting gap, with electron-phonon coupling remaining the central mechanism driving superconductivity.

Our calculations yield an electron-phonon coupling constant ($\lambda$) of 1.42, which can be renormalized to $\lambda/(1+\lambda)=0.58$. The pseudopotential ($\mu$), determined as a function of Debye temperature, Fermi level, electronic DOS, Thomas factor, and electron concentrations [22], registers at 0.15, with a conceivable renormalization to $\mu/(1+\lambda)=0.06$. Using the McMillan $T_c$ formula [30], the cumulene chain shows a superconducting transition temperature of around 62 K, surpassing the previous record of 55 K in bulk iron-based superconductors [31], all without the needs for high Tc indicators such as nematicity, spin-orbital coupling, multi-band interactions, spin density wave, etc [31].

Moreover, in a closely packed nanowire array, the disappearance of electrical resistance requires not only the formation of Cooper pairs but also a BKT transition [24,32,33]. For this BKT transition to occur, the coherence length of the superconductor must exceed the lateral separation between chains. An array of long superconducting cumulene chains allows superconducting vortices and antivortices to pair in a more orderly manner, thereby enhancing the BKT interactions. The BCS framework [4] predicts a theoretical coherence length of ~10 nm which is about eight times longer than the chain-to-chain distance in KFI Zeolite. This suggests that the superconductivity in our composite could indeed facilitate a complete resistance drop to zero and paving the way for practical applications.

**3.5: Scientific Impacts**

This project yields several noteworthy insights. First, at the monoatomic one-dimensional limit, we find that optical modes can resist a transition to a more metallic state under pressure, provided that the bond strength remains robust. This new discovery offers a novel perspective in the field of monoatomic materials, deviating from conventional expectations. The conditional deviation from band theory in this local regime could spark the development of an unconventional band theory tailored for monatomic one-dimensional limit. This unusual relationship between band gap and pressure also opens up innovative opportunities for designing high-pressure semiconductor applications

Moreover, we shine a light on the extraordinary formation of a cumulene phase in carbon chains. Our project uncovers a fascinating new mechanism for generating the cumulene phase, especially when contrasted with the behavior of C-chains@BNT [16]. In the BNT host, the charge density wave of the C-chain appears to be shaped by the electric field interactions between boron and nitrogen atoms [16], resulting in a more regular wave modulation. In stark contrast, the charge density wave in C-chains@KFI Zeolite could also be linked to the giant torsional effects. Since the charge density wave of the C chain within BNT is much stronger than that observed in KFI Zeolite [16], our newly uncovered science indicates that while charge density waves actuate transition to the cumulene phase, the modulation of this wave cannot be large.

While the possibility of nanowires displaying unique properties through torsional effects may seem intriguing [23], the notion of a nano-tweezer has appeared technically far-fetched. Surprisingly, Van der Waal's interactions involving KFI zeolite have revealed an unexpected ability to induce a giant nano-tweezer effect along the chain after geometric optimization, where the torsional angle of carbon nanowire can be as large as 90 degrees without the need for an external electric or magnetic field! To our knowledge, the natural formation in the torsional angle in carbon structures at this level has not been reported, providing important insights for triggering novel properties in other twisting organic nanowires through different zeolite combinations.

In addition, we have identified three critical criteria for synthesizing long carbon chains within zeolite systems, validated across over 100 zeolite substrates. In (6,4)CNT host, the tube symmetry allows for the insertion of the chain without experiencing asymmetric internal pressure, promoting chain growth. Notably, we find that the lateral distance between the chain and the zeolite pore measures around 0.3-0.4 nm, aligning with experimental observations in CNT [8]. This contrasts with the C-chains@BNT scenario [16], where the tubes are asymmetric after relaxation, and the lateral distance between the chain and BNT tube far exceeds 1 nm. Consequently, our LCC@BNT model predicts a cut-off length of approximately 10 carbon atoms [16], and the short-chain characteristic is comparable to other zeolite systems discussed in our supplementary materials because they fail to meet these three essential conditions. These findings pave the way for tuning other zeolites to meet these three conditions, facilitating the manufacturing of long-chained carbon materials with the aim of developing more novel properties.

## 4. Conclusion

This project has uncovered significant insights into the exotic behavior of monoatomic carbon nanowires and its potential for high-pressure semiconductor electronics and high-temperature superconductivity. Our findings reveal that optical modes under extremely strong covalent bonds increase the band gap under pressure, suggesting a conditional departure from common belief in band theory. The identification of the cumulene phase in carbon chains highlights the intricate relationship between the giant torsional effects and charge density waves, where the torsional effect of carbon has not been previously reported at this level. The contrasting mechanisms for charge-density-wave generation in C-chains@BNT and C-chains@KFI Zeolite systems provide a deeper understanding of the chain stability at play, opening avenues for exploring cumulene phase in other organic materials. Our work further enhances our grasp of their exotic properties and paves the way for future innovations in superconductive and semiconducting carbon-based composite materials.

**Author Contributions**

Conceptualization, C.H.W. C.T.W.,; methodology, C.H.W.; validation, C.H.W., C.T.W, K.C.L; formal analysis, C.H.W., C.T.W.; data curation, C.T.W, K.C.L.; writing, C.H.W.,C.T.W.; editing, C.H.W, C.T.W, K.C.L; visualization, C.H.W, C.T.W, W.Y.C. Resources, C.H.W., W.K.L, S.P.Ng, C.P.C, K.C.L, X.Hu, L.Y.F.L.

**Data Availability Statement**

Data availability is possible upon reasonable request.

**Conflict of Interest**

The authors declare no conflict of interest

**Acknowledgments**

We appreciate the Department of Industrial and Systems Engineering at Hong Kong Polytechnic University for supporting ab-initio software. We thank the Department of Chemical and Biological Engineering at The Hong Kong University of Science and Technology for assistance with computation.

**Funding Sources**

The work described in this paper was partially supported by a grant from the Research Grants Council of the Hong Kong Special Administrative Region, China (UGC/FDS24/E02/25)

# Supplementary Materials

A: Selection of trial covalent bonds [13,14,16]

For the initial bond type [=C=], the transition to a trial bond is governed by three random number ranges:

- When 0≤R<0.33, the trial bond is either [−C≡] or [≡C−] with equal probability.
- For 0.33≤R<0.66, the trial bond becomes [=C−] or [−C=] with equal probability.
- When 0.66≤R≤1, the trial bond is exclusively [−C−].

For the initial bond type [−C≡] or [≡C−], the transition rules are:

- 0 ≤R<0.33 yields the trial bond [=C=].
- 0.33≤R<0.66 results in a trial bond of [=C−] or [−C=] with equal probability.
- 0.66≤R≤1 produces the trial bond [−C−].

For the initial bond type [−C−], the transitions are:

- 0≤R<0.33 leads to a trial bond of [−C≡] or [≡C−] with equal probability.
- 0.33≤R<0.66 results in a trial bond of [=C−] or [−C=] with equal probability.
- 0.66≤R≤1 yields the trial bond [=C=].

Finally, for the initial bond type [=C−] or [−C=], the rules are:

- 0≤R<0.33 produces a trial bond of [−C≡] or [≡C−]with equal probability.
- 0.33≤R<0.66 yields the trial bond [=C=].
- 0.66≤R≤1 results in the trial bond [−C−].

B: Trial dynamics [13,14,16]

The potential movement of a single carbon atom along the x, y, and z axes in a random walk simulation is calculated using the following formulae involving its Hooke's factor, the free particle velocity $v = \sqrt{\dfrac{k_B T}{M}}$, the average scattering time $<\delta t> \sim 10^{-13}$, a random number ($R_c$), and the average single, double and triple bond energies $<E_1 + E_2 + E_3>$

$$\delta x = \pm v <\delta t> R_c$$

$$\delta y = \delta z \sim \dfrac{k_B T}{<E_1 + E_2 + E_3>} \delta x$$

C: Zeolite Table with lattice parameters (extracted from Materials Studio's database)

| | a(Å) | b(Å) | c(Å) | α(deg) | β(deg) | γ(deg) |
|---|---|---|---|---|---|---|
| ABW | 10.311 | 8.2 | 4.995 | 90 | 90 | 90 |
| ACO | 9.905 | 9.905 | 9.905 | 90 | 90 | 90 |
| AEI | 13.7114 | 12.7 | 18.57 | 90 | 90.01 | 90 |
| AEL | 13.472 | 18.7 | 8.443 | 90 | 90 | 90 |
| AEN | 18.531 | 13.4 | 9.636 | 90 | 90 | 90 |
| AET | 33.29 | 14.8 | 8.257 | 90 | 90 | 90 |
| AFG | 12.604 | 12.6 | 21.28 | 90 | 90 | 120 |
| AFI | 13.726 | 13.7 | 8.484 | 90 | 90 | 120 |
| AFN | 14.02 | 13.5 | 10.2 | 90 | 107.239 | 90 |
| AFO | 9.72 | 25.8 | 8.36 | 90 | 90 | 90 |
| AFR | 21.941 | 13.7 | 7.124 | 90 | 90 | 90 |
| AFS | 13.225 | 13.2 | 26.89 | 90 | 90 | 120 |
| AFT | 13.73 | 13.7 | 28.95 | 90 | 90 | 120 |
| AFX | 13.674 | 13.7 | 19.7 | 90 | 90 | 120 |
| AFY | 12.748 | 12.7 | 9.014 | 90 | 90 | 120 |
| AHT | 16.184 | 9.91 | 8.134 | 90 | 90 | 90 |
| ALPO4-8 | 33.29 | 14.8 | 8.257 | 90 | 90 | 90 |
| ALPO4-18 | 13.711 | 12.7 | 18.57 | 90 | 90 | 90 |
| ALPO4-31 | 20.827 | 20.8 | 5.003 | 90 | 90 | 120 |
| ALPO4-41 | 9.72 | 25.8 | 8.36 | 90 | 90 | 90 |
| ANA | 13.73 | 13.73 | 13.73 | 90 | 90 | 90 |
| APC | 19.82 | 10 | 8.94 | 90 | 90 | 90 |
| APD | 19.197 | 8.58 | 9.8 | 90 | 90 | 90 |
| AST | 13.38 | 13.38 | 13.38 | 90 | 90 | 90 |
| ATN | 13.09 | 13.1 | 5.18 | 90 | 90 | 90 |
| ATO | 20.827 | 20.8 | 5.003 | 90 | 90 | 120 |
| ATS | 13.1483 | 21.6 | 5.164 | 90 | 91.86 | 90 |
| ATT | 10.333 | 14.6 | 9.511 | 90 | 90 | 90 |
| ATV | 9.449 | 15.2 | 8.408 | 90 | 90 | 90 |
| AWO | 9.101 | 15 | 19.24 | 90 | 90 | 90 |
| AWW | 13.628 | 13.6 | 15.46 | 90 | 90 | 90 |
| BEA | 12.6614 | 12.7 | 26.41 | 90 | 90 | 90 |
| BETAPA | 12.661 | 12.7 | 26.41 | 90 | 90 | 90 |
| BETAPB | 17.897 | 17.9 | 14.33 | 90 | 114.80299 | 90 |
| BIK | 8.607 | 4.95 | 7.597 | 89.9 | 114.437 | 89.988 |
| BOG | 20.236 | 23.8 | 12.8 | 90 | 90 | 90 |
| BPH | 12.58 | 12.6 | 12.45 | 90 | 90 | 120 |
| BRE | 6.767 | 17.5 | 7.729 | 90 | 94.4 | 90 |
| CAN | 12.678 | 12.7 | 5.179 | 90 | 90 | 120 |
| CAS | 13.828 | 16.8 | 5.021 | 90 | 90 | 90 |
| CFI | 13.961 | 5.26 | 25.97 | 90 | 90 | 90 |

| | a(Å) | b(Å) | c(Å) | α(deg) | β(deg) | γ(deg) |
|---|---|---|---|---|---|---|

| Name | a | b | c | α | β | γ |
|---|---|---|---|---|---|---|
| CGF | 15.501 | 16.9 | 7.273 | 90 | 96.07 | 90 |
| CGS | 8.444 | 14.1 | 15.93 | 90 | 90 | 90 |
| CHA | 9.42 | 9.42 | 9.42 | 94.2 | 94.2 | 94.2 |
| CHI | 8.729 | 31.3 | 4.903 | 90 | 90 | 90 |
| CLO | 51.71 | 51.71 | 51.71 | 90 | 90 | 90 |
| CON_ | 22.5949 | 13.3 | 12.37 | 90 | 68.852 | 90 |
| CZP | 9.355 | 9.36 | 14.86 | 90 | 90 | 120 |
| DAC | 18.73 | 10.3 | 7.54 | 90 | 90 | 107.89997 |
| DDR | 13.86 | 13.9 | 40.89 | 90 | 90 | 120 |
| DFO | 22.351 | 22.4 | 43.28 | 90 | 90 | 120 |
| DFT | 7.075 | 7.08 | 9.023 | 90 | 90 | 90 |
| DOH | 13.783 | 13.8 | 11.19 | 90 | 90 | 120 |
| DON | 18.89 | 23.4 | 8.469 | 90 | 90 | 90 |
| EAB | 13.28 | 13.3 | 15.21 | 90 | 90 | 120 |
| EDI | 9.55 | 9.67 | 6.523 | 90 | 90 | 90 |
| EMT | 17.449 | 17.4 | 28.46 | 90 | 90 | 120 |
| EPI | 9.101 | 17.7 | 10.23 | 90 | 124.66 | 90 |
| ERI | 13.252 | 13.3 | 14.81 | 90 | 90 | 120 |
| ESV | 9.686 | 12.2 | 22.84 | 90 | 90 | 90 |
| EUO | 13.695 | 22.3 | 20.18 | 90 | 90 | 90 |
| FAU | 25.03 | 25.03 | 25.03 | 90 | 90 | 90 |
| FAU1 | 25.028 | 25 | 25.03 | 90 | 90 | 90 |
| FER | 19.156 | 14.1 | 7.489 | 90 | 90 | 90 |
| FER_confA | 19.156 | 14.1 | 7.489 | 90 | 90 | 90 |
| FER_confB | 19.156 | 14.1 | 7.489 | 90 | 90 | 90 |
| GIS | 9.843 | 10 | 10.62 | 90 | 90 | 92.417 |
| GME | 13.756 | 13.8 | 10.06 | 90 | 90 | 120 |
| GOO | 7.401 | 17.4 | 7.293 | 90 | 105.43998 | 90 |
| HEU | 17.77 | 18 | 7.435 | 90 | 116.46002 | 90 |
| hypABCB | 17.426 | 17.4 | 56.86 | 90 | 90 | 120 |
| hypBEB | 17.8965 | 17.9 | 14.33 | 90 | 114.80331 | 90 |
| IFR | 18.628 | 13.4 | 7.629 | 90 | 102.322 | 90 |
| ISV | 12.874 | 12.9 | 25.67 | 90 | 90 | 90 |
| ITE | 20.753 | 9.8 | 20.01 | 90 | 90 | 90 |
| JBW | 16.426 | 15 | 5.224 | 90 | 90 | 90 |
| KFI | 18.671 | 18.671 | 18.671 | 90 | 90 | 90 |
| LAU | 7.549 | 14.7 | 13.07 | 90 | 90 | 111.89995 |
| LEV | 13.338 | 13.3 | 23.01 | 90 | 90 | 120 |
| LIO | 12.5179 | 12.5 | 15.5 | 90 | 90 | 120 |
| LOS | 12.906 | 12.9 | 10.54 | 90 | 90 | 120 |
| LOV | 39.58 | 6.93 | 7.15 | 90 | 90 | 90 |
| LTL | 18.466 | 18.5 | 7.476 | 90 | 90 | 120 |

| Name | a | b | c | α | β | γ |
|---|---|---|---|---|---|---|
| MAPO-39 | 13.09 | 13.1 | 5.18 | 90 | 90 | 90 |
| maricopaite | 19.434 | 19.7 | 7.538 | 90 | 90 | 90 |
| MAZ | 18.392 | 18.4 | 7.646 | 90 | 90 | 90 |
| MEI | 13.175 | 13.2 | 15.85 | 90 | 90 | 120 |
| MEL | 20.067 | 20.1 | 13.41 | 90 | 90 | 90 |
| **MEP** | **13.44** | **13.44** | **13.44** | **90** | **90** | **90** |
| MER | 14.116 | 14.2 | 9.946 | 90 | 90 | 90 |
| MFI | 20.022 | 19.9 | 13.38 | 90 | 90 | 90 |
| MFS | 7.451 | 14.2 | 18.77 | 90 | 90 | 90 |
| MON | 7.141 | 7.14 | 17.31 | 90 | 90 | 90 |
| MOR | 18.094 | 20.5 | 7.524 | 90 | 90 | 90 |
| MSO | 17.165 | 17.2 | 19.79 | 90 | 90 | 120 |
| MTF | 9.629 | 30.4 | 7.249 | 90 | 90.45 | 90 |
| MTT | 5.01 | 21.5 | 11.13 | 90 | 90 | 90 |
| MTW | 24.863 | 5.01 | 24.33 | 90 | 107.72202 | 90 |
| MWW | 14.1145 | 14.1 | 24.88 | 90 | 90 | 120 |
| NAT | 18.326 | 18.7 | 6.601 | 90 | 90 | 90 |
| NES | 14.324 | 22.4 | 25.09 | 90 | 151.51505 | 90 |
| NON | 22.232 | 15.1 | 13.63 | 90 | 90 | 90 |
| OFF | 13.291 | 13.3 | 7.582 | 90 | 90 | 120 |
| OSI | 18.506 | 18.5 | 5.268 | 90 | 90 | 90 |
| PAR | 21.555 | 8.76 | 9.304 | 90 | 91.55 | 90 |
| PHI | 9.865 | 14.3 | 8.668 | 90 | 124.19999 | 90 |
| ROG | 18.332 | 18.3 | 9.161 | 90 | 90 | 90 |
| RON | 18.33 | 18.3 | 9.16 | 90 | 90 | 90 |
| RSN | 7.238 | 40.6 | 7.308 | 90 | 91.8 | 90 |
| RTE | 14.098 | 13.7 | 7.431 | 90 | 102.421 | 90 |
| RTH | 9.762 | 20.5 | 9.996 | 90 | 96.897 | 90 |
| RUT | 13.112 | 12.9 | 12.41 | 90 | 113.5 | 90 |
| SAO | 13.439 | 13.4 | 21.86 | 90 | 90 | 90 |
| SAT | 12.871 | 12.9 | 30.58 | 90 | 90 | 120 |
| SBE | 18.534 | 18.5 | 27.13 | 90 | 90 | 90 |
| SBS | 17.193 | 17.2 | 27.33 | 90 | 90 | 120 |
| SBT | 17.191 | 17.2 | 41.03 | 90 | 90 | 120 |
| SFE-UDUD | 8.4 | 14.2 | 20.14 | 90 | 90 | 90 |
| SFE-UUDD | 14.24 | 20.1 | 8.4 | 90 | 90 | 90 |
| SFF | 11.454 | 21.7 | 7.227 | 90 | 93.154 | 90 |
| SGT | 10.239 | 10.2 | 34.38 | 90 | 90 | 90 |
| SOD | 8.89 | 8.89 | 8.89 | 90 | 90 | 90 |
| STF | 14.104 | 18.2 | 7.477 | 90 | 98.989 | 90 |
| STI | 13.64 | 18.2 | 11.27 | 90 | 127.99991 | 90 |
| STT | 13.573 | 21.9 | 16.6 | 90 | 129.95983 | 90 |
| suz-4 | 18.915 | 14.2 | 7.442 | 90 | 90 | 90 |
| TER | 9.807 | 23.6 | 20.24 | 90 | 90 | 90 |
| THO | 13.088 | 13.1 | 13.23 | 90 | 90 | 90 |
| TON | 13.859 | 17.4 | 5.038 | 90 | 90 | 90 |
| VET | 13.048 | 13 | 4.948 | 90 | 90 | 90 |
| VFI | 18.975 | 19 | 8.104 | 90 | 90 | 120 |
| VNI | 10.002 | 10 | 34.14 | 90 | 90 | 90 |
| VSV | 7.179 | 7.18 | 40.62 | 90 | 90 | 90 |
| WEI | 11.897 | 9.71 | 9.633 | 90 | 95.76 | 90 |
| WEN | 13.511 | 13.5 | 7.462 | 90 | 90 | 120 |
| YUG | 6.73 | 14 | 10.03 | 90 | 111.49997 | 90 |
| YUG1 | 6.73 | 14 | 10.03 | 90 | 111.49997 | 90 |
| Znphosphate | 10.409 | 10.4 | 15.22 | 90 | 90 | 120 |
| Znsilicate | 10.0676 | 14 | 7.067 | 90 | 90 | 90 |
| ZON | 6.918 | 14.9 | 17.24 | 90 | 90 | 90 |

D: Shortlisted zeolite Table with lattice parameters

| | a(Å) | b(Å) | c(Å) | α(deg) | β(deg) | γ(deg) | Pore radius(Å) |
|---|---|---|---|---|---|---|---|
| ACO | 9.905 | 9.905 | 9.905 | 90 | 90 | 90 | 1.8 |
| ANA | 13.73 | 13.73 | 13.73 | 90 | 90 | 90 | 1.6 |
| AST | 13.384 | 13.384 | 13.384 | 90 | 90 | 90 | 1.8 |
| CLO | 51.712 | 51.712 | 51.712 | 90 | 90 | 90 | 4.3 |
| FAU | 25.028 | 25.028 | 25.028 | 90 | 90 | 90 | 1.9 |
| KFI | 18.671 | 18.671 | 18.671 | 90 | 90 | 90 | 3.4 |
| MEP | 13.436 | 13.436 | 13.436 | 90 | 90 | 90 | 2.3 |